\RequirePackage{fix-cm}
\documentclass[smallextended]{svjour3}       
\smartqed  
\usepackage{graphicx,natbib,url}
\usepackage{aas_macros}
\usepackage{xcolor}

\begin{document}

\title{CAPTURE: A continuum imaging pipeline for the uGMRT
}
\subtitle{}


\author{Ruta Kale        \and
        C. H. Ishwara-Chandra  
}


\institute{Ruta Kale \at
              National Centre for Radio Astrophysics, Tata Institute of Fundamental Research, S. P. Pune University Campus, Ganeshkhind, Pune 411007, India \\             
              \email{ruta@ncra.tifr.res.in}           
           \and
           C. H. Ishwara-Chandra \at
              National Centre for Radio Astrophysics, Tata Institute of Fundamental Research, S. P. Pune University Campus, Ganeshkhind, Pune 411007, India
}

\date{Received: date / Accepted: date}

\maketitle

\begin{abstract}
We present the first fully automated pipeline for making images from the interferometric data obtained from the upgraded Giant Metrewave Radio Telescope (uGMRT) 
called CAsa Pipeline-cum-Toolkit for Upgraded Giant Metrewave Radio Telescope data REduction - CAPTURE. 
It is a python program that uses tasks from the NRAO Common Astronomy Software Applications (CASA) to perform the steps of flagging of bad data, calibration, 
imaging and self-calibration.  The salient features of the pipeline are: i) a fully automatic mode to go from the raw data to a self-calibrated 
continuum image, ii) specialized flagging strategies for short and long baselines that ensure minimal loss of extended structure, iii) flagging of persistent narrow 
band radio frequency interference (RFI), iv) flexibility for the user to configure the pipeline for step-by-step analysis or special cases and 
v) analysis of data from the legacy GMRT. CAPTURE is available publicly on github (\url{https://github.com/ruta-k/uGMRT-pipeline}, release v1.0.0). 
The primary beam correction for the uGMRT images produced with CAPTURE is made separately available 
 at \url{https://github.com/ruta-k/uGMRTprimarybeam}. We show examples of using CAPTURE on uGMRT and legacy GMRT data. In principle, CAPTURE can be tailored for use 
 with other radio interferometric data.


\end{abstract}

\section{Introduction}
Radio interferometric data require post-processing that involves 
calibration and inverse Fourier transform to obtain an image of the sky. 
The Giant Metrewave Radio Telescope (GMRT), located near Pune, India, is a radio interferometer 
 that consists of 30 antennas of 45 m diameter each{, with the longest baseline $\sim25$ km}, 
 that recently underwent an upgrade of the receivers. 
 Before the upgrade it operated in 5 frequency bands in the range of 120 - 1450 MHz with instantaneous bandwidths of 33.3 MHz. 
 The upgrade included broadbanding of receivers and backend which now allows recording data with instantaneous 
 bandwidths of up to 400 MHz \citep{Gupta2017}. We will refer to the older GMRT with narrow band 
 receivers as the ``legacy GMRT'' and the one after the upgrade as the upgraded GMRT (uGMRT). 
 The uGMRT is currently operational in four frequency bands, namely, 120 - 250 MHz (band-2), 250 - 500 MHz (band-3), 
 550 - 850 MHz (band-4) and 1000 - 1460 MHz (band-5).

The standard data analysis software for the legacy GMRT was the NRAO Astronomical Image Processing 
System (AIPS\footnote{\url{http://www.aips.nrao.edu/index.shtml#WHATISIT}}). 
An AIPS based continuum imaging pipeline, Source Peeling and Atmospheric Modelling (SPAM) \citep{2014ASInC..13..469I} 
has been available mainly for the lower frequency bands ($< \mathrm{GHz}$) of the GMRT \citep{tgssadr}. 

Analysis of the uGMRT wideband data has been carried out using semi-automated methods so far \citep[e. g.][]{2018MNRAS.480.5352K,2019arXiv191108904R} and has been challenging.
The fractional bandwidths of uGMRT, $\Delta \nu/\nu \sim 0.3 - 0.7 $ (band-5 to band-2) require imaging algorithms that simultaneously account for the wideband and 
widefield effects which have not been implemented in AIPS. While the typical datasets for a few hours observing duration under default settings (16s sampling time and 
 256 frequency channels) were of a few GB in sizes for the legacy GMRT, these are typically of several tens of GB for 
 the uGMRT. The wide banding also meant increased radio frequency interference (RFI) particularly at low radio frequencies, which
along with the large data size has made the manual flagging and processing of the data highly challenging and impractical.
 Currently there is no publicly available pipeline for the analysis of uGMRT wideband data.

The Common Astronomy Software Applications (CASA) package \citep{2007ASPC..376..127M} has algorithms for widefield and wideband imaging of the
interferometric data such as the w-projection \citep{2005ASPC..347...86C} and multi-scale multi-frequency synthesis \citep{2011A&A...532A..71R} implemented.
In this paper we present, CAsa Pipeline-cum-Toolkit for Upgraded Giant Metrewave Radio Telescope data REduction (CAPTURE), 
which is a python program that uses CASA to produce continuum images from radio interferometric data.
The data reduction is carried out in the standard steps of flagging (removal of bad data), calibration, imaging 
and self-calibration. It is possible to start from the rawdata obtained from the observatory in the native lta format 
and obtain a self-calibrated continuum image using CAPTURE in a fully automated manner. CAPTURE can also be used like a tool-kit to carry out a 
step by step analysis. In addition it can be adapted for special imaging requirements.

The paper is organised as follows. The structure of CAPTURE is described in Sec. ~\ref{structure}. The usage of it is presented 
in Sec. ~\ref{usage} and the examples in Sec. ~\ref{eg}. The method for primary beam correction is described in Sec.~\ref{pb}. 
Special cases for running CAPTURE are described in Sec.~\ref{spcases} and the current limitations are summarised in Sec.~\ref{limits}. 
Conclusions are presented in Sec.~\ref{conclu}.

\section{The structure of CAPTURE}\label{structure}
\begin{figure}
    \centering
        \includegraphics[height=15cm]{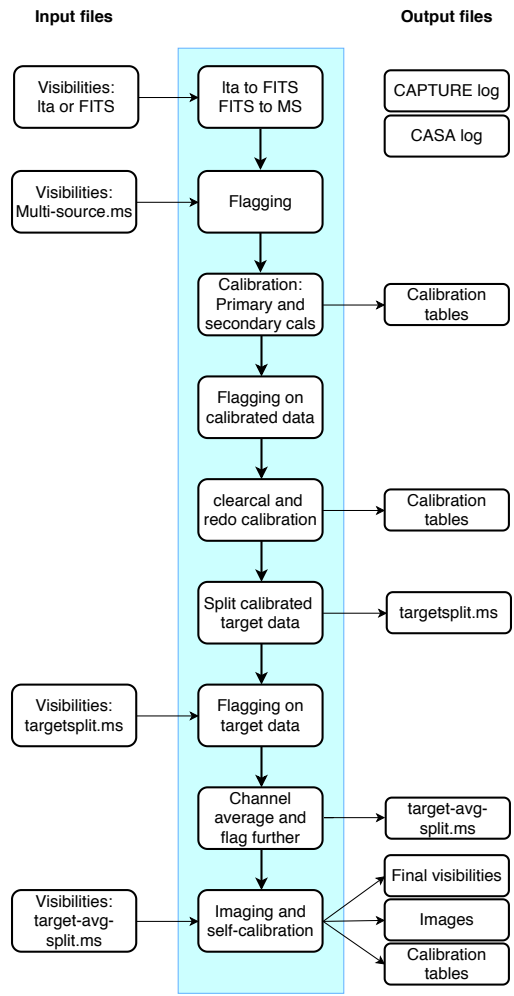}
    \caption{{A schematic diagram showing the pipeline processes in the central block and the input and output files are shown on the left and right, 
    respectively.} The shaded rectangle on the top shows the block where the the multi-source measurement set is being processed. 
    The lower shaded block is where the split file with a single target source is processed. By appropriately choosing the configuration, 
    it is possible to stop and examine the outputs at every block in both the shaded regions.}
    \label{fig:flowchart}
\end{figure}

The inputs for CAPTURE are the visibilities in either of lta, FITS or Measurement Set (MS) format and the outputs are calibration tables, 
calibrated target visibilities and self-calibrated images. The operations carried out by CAPTURE are schematically shown in Fig.~\ref{fig:flowchart}. 
Once the MS file of the visibilities is available, the analysis steps in CAPTURE begin. The first part is to locate 
and flag non-working antennas based on the calibrator scans. If a calibrator scan is found to be bad then the 
adjoining source scans are also flagged. After the initial flagging, the data are calibrated and a round of flagging is done on 
the calibrated data. With the improved flagging the calibration is carried out again after clearing the earlier calibration. 
These calibrated data of the target are 
split into a separate file and inspected for further bad data. After this round of flagging, the data are averaged in frequency according to users' preference 
and again inspected for bad data. In the use of automated flagging modes tfcrop and rflag in the task flagdata for the target source, 
baselines with the central square antennas (C00, ..., C14) are flagged with more conservative thresholds than the arm antennas 
in order to prevent the loss of extended struture due to over-flagging.
These data are then taken up for imaging and self-calibration. During self-calibration, flagging is carried out on the 
residual column data to remove the low lying RFI. 

CAPTURE can carry out an end to end processing of raw data to self-calibrated images once the inputs are set by the user. 
If there is uncertainty regarding the inputs for imaging or if the user feels the need to carry out manual flagging 
at intermediate stages after inspection of the data, the pipeline has the flexibility to run in stages on the data.

The github repository for CAPTURE contains the files listed in Table~\ref{gitfiles}.

\begin{table}
\caption{The CAPTURE repository.}\label{gitfiles}
\begin{tabular}{ll}
\hline
  File & Description \\
  \hline
config\_capture.ini & Configuration file.\\
capture.py & Code for processing.\\
ugfunctions.py & Python functions.\\
vla-cals.list & List of calibrator names\\
listscan, gvfits & Precompiled binaries for lta to FITS conversion.\\
\hline
\end{tabular}

 
\end{table}

\subsection{Configuration}\label{initblock}
The user will set the inputs for CAPTURE in the file named config\_capture.ini. The first part of the input parameters  
take Boolean values (True / False) and will set the state for the processing of the data. The second part is 
where the filenames and other details for calibration and imaging are set. Both of these parts are under the ``basic'' 
block in the configuration file.
The list of parameters in this block is provided in Table~\ref{initsetupnew} with a brief explanation for each of them.
In addition to this a ``default'' block is defined in the this file and the parameters are given in Table~\ref{initsetupnew-default}. 
The parameters in this part are to be left unchanged for the normal modes of running the pipeline.

\begin{table}
\centering
\caption{Parameters in the ``basic'' block of config\_capture.ini (Sec.~\ref{initblock}).}\label{initsetupnew}
 \begin{tabular}{lp{0.5cm}p{2.0cm}p{5.5cm}l}
 \hline
 Parameter      &     & Value   & Description\\
 \hline
fromlta         & $=$ & True    & Convert lta file to FITS.              \\
fromfits        & $=$ & True    & Convert FITS to MS.              \\
frommultisrcms  & $=$ & True    & When working with a multi-source MS file.              \\
findbadants     & $=$ & True    & Find bad antennas.              \\
flagbadants     & $=$ & True    & Find and flag bad antennas.              \\
findbadchans    & $=$ & True    &  Find bad channels within known RFI affected frequency ranges.             \\
flagbadfreq     & $=$ & True    &  Find and flag bad channels.             \\
flaginit        & $=$ & True    &  Do initial flagging.             \\
doinitcal       & $=$ & True    &  Calibrate the data.           \\
doflag          & $=$ & True    &   Flag on the calibrated data.            \\
redocal         & $=$ & True    &   Re-do the calibration.            \\
dosplit         & $=$ & True    &   Split the target source data.            \\
flagsplitfile   & $=$ & True    &   Flag the target source data.            \\
dosplitavg      & $=$ & True    &   Average target source data in frequency.            \\
doflagavg       & $=$ & True    &   Flag on the frequency averaged data.            \\
makedirty       & $=$ & True    &   Make a dirty image.            \\
doselfcal       & $=$ & True    &   Run imaging and self-calibration.            \\
\hline
ltafile         & $=$ & file.lta         &  Name of the lta file.             \\
gvbinpath       & $=$ & ./listscan,./gvfits     &  Path to the listscan and gvfits executables.             \\
fits\_file       & $=$ & TEST.FITS        &   Name of the FITS file.            \\
msfilename      & $=$ & test.ms      &   Name of the multi-source MS file.            \\
splitfilename   & $=$ &     &   Name of the split file if available.            \\
splitavgfilename   & $=$ &          & Name of the frequency averaged file if available.              \\
setquackinterval   & $=$ &      10      &  Time in seconds to flag at the beginning and at the end of each scan.             \\
ref\_ant         & $=$ & C00     &    Provide reference antenna name. Note that the antenna number will not work.           \\
clipfluxcal     & $=$ & 0.0,80.0        &   Clipminmax levels in Jy for flux calibrator.            \\
clipphasecal    & $=$ & 0.0,60.0        &  Clipminmax levels in Jy for the secondary calibrator/s.             \\
cliptarget      & $=$ & 0.0,30.0        &  Clipminmax in Jy for target source.             \\
clipresid       & $=$ & 0.0,10.0        &  Clipminmax in Jy for residual column used only during self-calibration.              \\
chanavg         & $=$ & 50      &   Number of channels to average. Choose in order to avoid bandwidth smearing.            \\
imcellsize      & $=$ & 1.0arcsec       &   Cell size for imaging.            \\
imsize\_pix      & $=$ & 7000     & Image size in pixel units.              \\
scaloops        & $=$ & 8       & Total number of self-calibration loops (including both phase-only and amplitude and phase).             \\
mJythreshold    & $=$ & 0.01    & A setting equal to the expected rms in mJy is found to work fine.              \\
pcaloops        & $=$ & 4       &  Number of phase-only self-calibration loops; should be $<=$ scaloops.             \\
scalsolints     & $=$ & 8.0min, 4.0min, & List of ``solint''s for the  \\
                &     & 2.0min, 1.0min,  &task gaincal in self-calibration. \\
                &     & 4.0min, 2.0min, 1.0min, 1.0min  & \\
niter\_start     & $=$ & 1000      & Number of iterations for the first imaging in the self-calibration loop. \\
use\_nterms      & $=$ & 2       & The nterms parameter used in tclean.               \\
nwprojpl        & $=$ & -1      & Number of w-projection planes; $-1$ implies that it is determined internally in tclean.              \\
\hline
\hline
\end{tabular}
 \end{table}
 
\begin{table}
\centering
\caption{Parameters in the ``default'' block of config\_capture.ini (Sec.~\ref{initblock}).}\label{initsetupnew-default}
 \begin{tabular}{lp{0.5cm}p{2.0cm}p{5cm}l}
 \hline
Parameter      &     & Value   & Description\\
 \hline
uvracal         & $=$ &         & UV-range cutoff used in calibration (not tested).              \\
uvrascal        & $=$ &         &  UV-range cutoff used in self-calibration (not tested).             \\
target          & $=$ & True    &  Always set to True.             \\
usetclean       & $=$ & True    &  Always set to True.             \\

\hline
 \end{tabular}
 \end{table}


\subsection{Functions}~\label{func}
The python functions that utilise CASA tasks for the operation of CAPTURE are defined in the file ``ugfunctions.py''. 
No modification to this file is expected for the standard operations on the data. Modifications to this part
for special cases of data analysis are discussed in Sec.~\ref{spcases}.

\subsection{Application}
This block carries out the actual processing of the files using the functions (Sec.~\ref{func}) with inputs provided by the user 
(Sec.~\ref{initblock}). 
{The output files are the calibration tables, measurement sets and image files in each iteration of self-calibration.} 
All the extension files generated by the 
task ``tclean'' will be retained. A FITS format file is created from the image in each iteration. The visibilities from 
the final self-calibration iteration are retained but those from the intermediate versions of self-calibration are removed. 
All the calibration tables produced in the initial and self-calibrations are retained. 
The lta, FITS and multi-source measurement are also retained.

\section{Using the pipeline}\label{usage}
{ 
CAPTURE is to be run with the command ``casa -c capture.py'' 
on the terminal in the same folder where the data are stored. CAPTURE will create all the new files in the same folder. 
Two log files will be created for each instance of running CAPTURE with the names ``casa-capture$\_$HH$\_$MM$\_$SS$\_$DD$\_$MM$\_$YYYY.log'' 
and ``capture$\_$HH$\_$MM$\_$SS$\_$DD$\_$MM$\_$YYYY.log''.
 The one with prefix ``casa'' contains the messages from the CASA logger and the other log file contains the reports from the pipeline processing. The log file will 
 also contain a summary of flagging on the visibilities.}

\subsection{Standard cases}\label{std}
The most common contents of an observing session are the scans on one or more standard flux calibrators, a secondary calibrator and 
a single target source. The user can start with the raw data file (the lta file) and obtain a self-calibrated image of the target 
source using CAPTURE by setting all the parameters to True in the initial set-up and providing all the inputs in the inputs block.
The cases of multiple primary and secondary calibrators for the same target source are implemented {in} CAPTURE. The special case where 
the primary calibrator also happens to be the secondary calibrator for the target is also handled in CAPTURE. 

In the current implementation the uv-distance limitations for the calibrators are not taken into account. The primary and secondary 
calibrators are identified by matching the names with their standard names in the J2000 IAU format as in the VLA calibrator manual \footnote{The 
file with this list vla-cals.list is provided in the git repository. The calibrator names are taken from \url{http://www.vla.nrao.edu/astro/calib/manual/csource.html}}. 
A source with a name different from that in the list is interpreted as a target source. More options for user controls in case of 
conflicts in source and calibrator names are given in Sec.~\ref{spcases}.

\subsection{Multiple targets}\label{mtar}
It is possible that in a single lta file, observations towards multiple targets with 
multiple phase calibrators are present. Since CAPTURE handles self-calibration of only one target 
at a time, the pipeline should be run in two steps by using the controls provided by the 
configuration file. In the first run, the Boolean parameters need to be set to True until the parameter ``dosplit'' - where the calibrated data on each target 
source is split into separate MS files and the subsequent ones to False. With this choice the data towards 
all the targets are calibrated and split into single source files.
It is recommended to create a separate folder for each source by the user and move the split files for the targets to the respective 
directories and make a copy of the CAPTURE code to it. In the second step, the initial set-up parameters from ``flagsplitfile'' onwards are set to True 
and all those before are set to False. The name of the target MS file should be 
provided in the input parameter ``splitfilename''. The imaging and self-calibration part will be run on this target. 
This can be repeated for all the targets in the respective directories. 

\subsection{Step by step analysis}\label{stepbystep}
If the user is unfamiliar with the data in terms of the observing frequency and 
the number of targets in the file, the configuration parameter with Boolean inputs allow the user to process one step at a time and check the output. 
The output of CASA task ``listobs'' and a list of bad antennas for each scan are generated which can be examined to choose the settings 
for calibration and imaging in the subsequent steps.

The initial set-up allows the user to halt CAPTURE after every major step { represented in the shaded part} of the 
schematic diagram shown in Fig.~\ref{fig:flowchart}. For example if the user wants to stop the pipeline after the recalibration to inspect the 
effect of the flags on the data, they can do so by setting the parameters from ``dosplit'' onwards to False and setting those before 
this to True. The user can also manually flag the data if preferred and start running the remaining part of the pipeline after 
the manual flagging is done. This mode of CAPTURE provides the flexibility to the user to control the pipeline effectively and 
also supplement it with additional manual analysis. 


\section{Examples}\label{eg}
We have used CAPTURE for making images using the legacy GMRT data and the  
uGMRT data at bands 3, 4 and 5. Here we show two examples of {the use of }CAPTURE pipeline: a uGMRT band-3 dataset 
and a 608 MHz legacy GMRT dataset. The details of the observations used are given in Table~\ref{obs}.

\begin{figure}
    \centering
        \includegraphics[trim = {4.0cm 6.0cm 3.0cm 6.0cm}, clip,height=12cm]{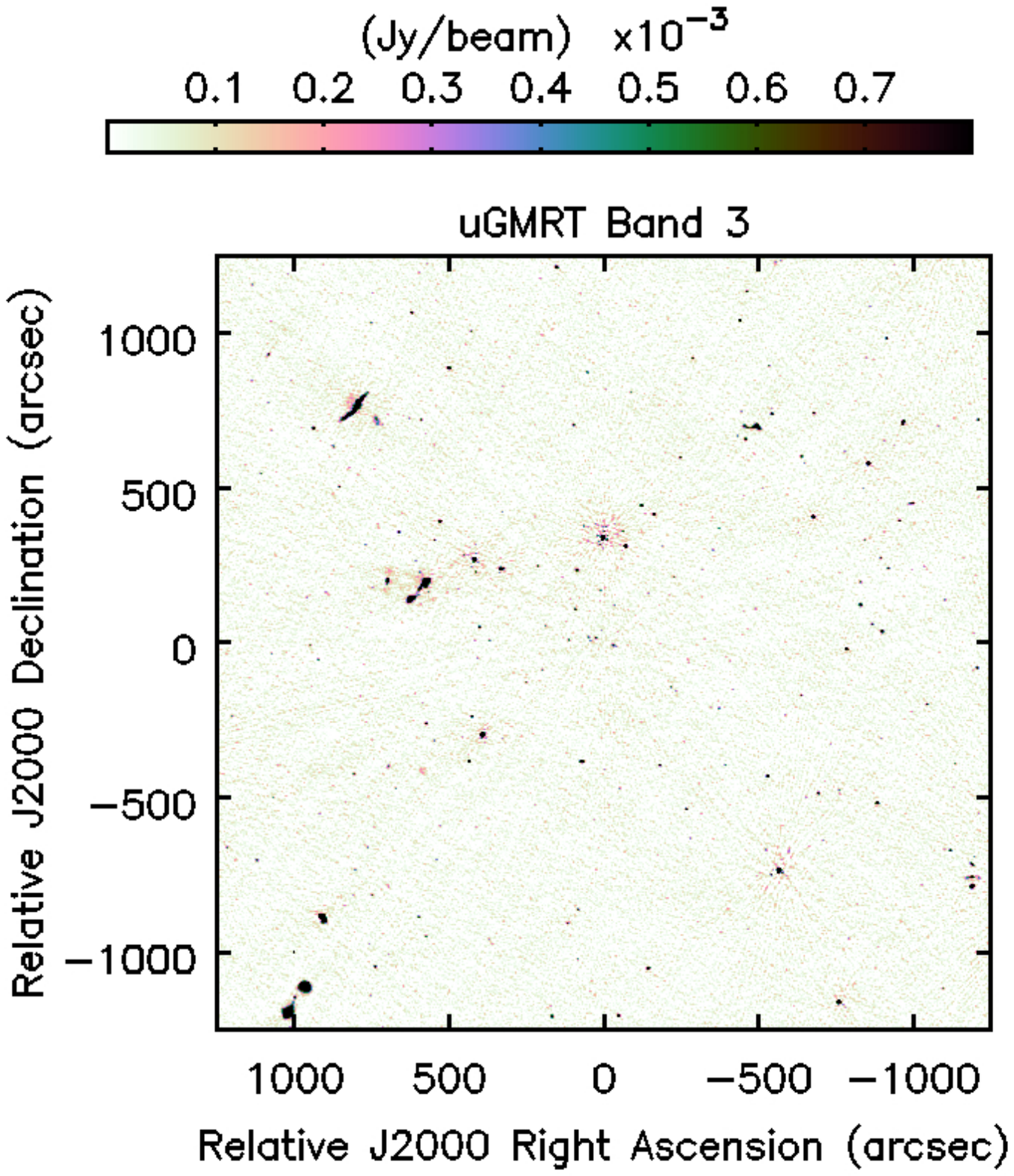}
    \caption{A uGMRT band-3 image with rms noise 43 $\mu$Jy beam$^{-1}$. The field is located at declination $\sim 0^{\circ}$.}
    \label{fig:ugmrt}
\end{figure}

\begin{figure}
    \centering
    \includegraphics[trim = {4.0cm 6.0cm 3.0cm 6.0cm}, clip,height=12cm]{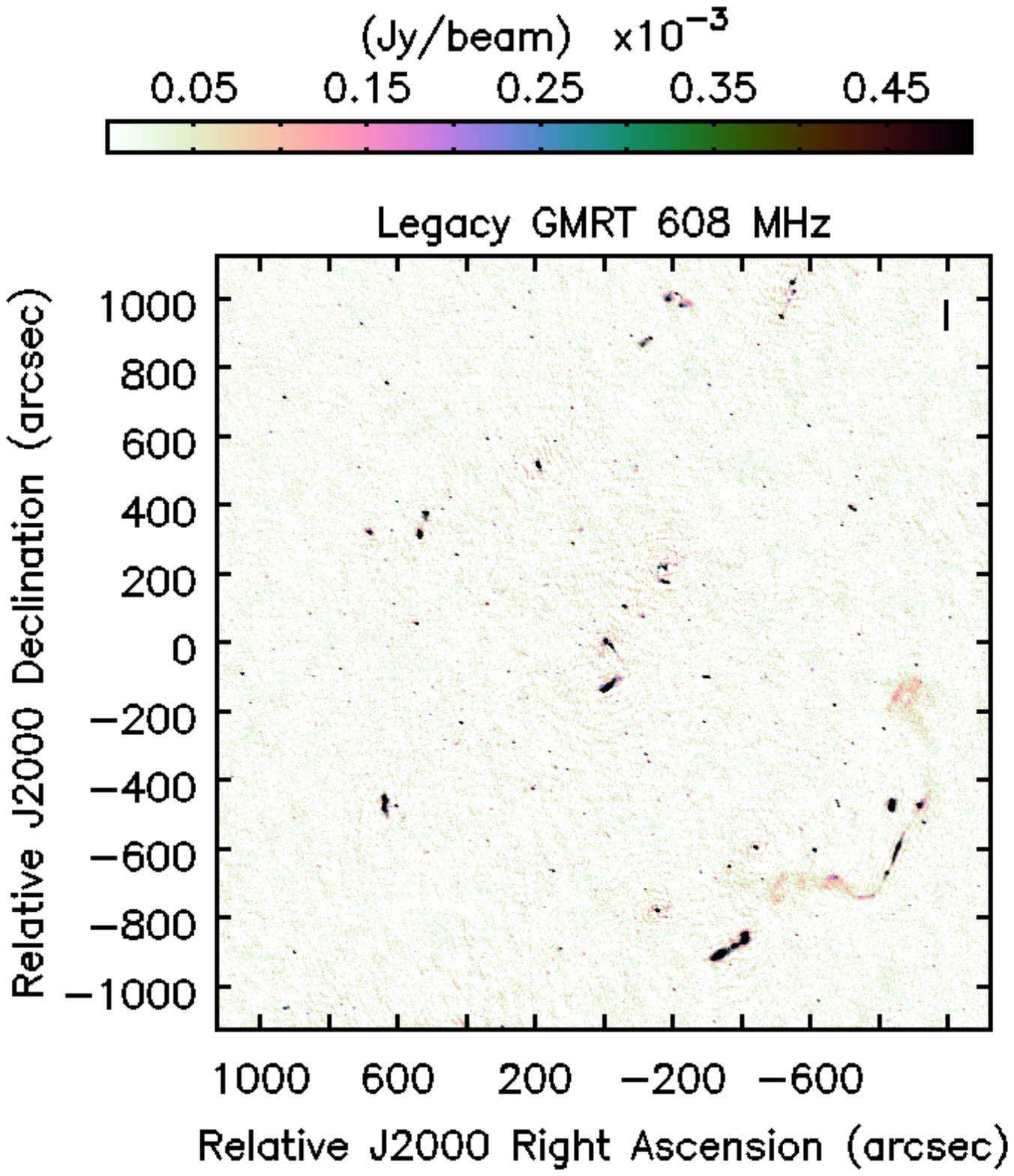}
    \caption{A legacy GMRT image at 608 MHz with rms 28 $\mu$Jy beam$^{-1}$. The declination of the field is $\sim +40^{\circ}$.}
    \label{fig:legacy}
\end{figure}

For the uGMRT dataset, CAPTURE was run in two stages as the case corresponded to that of a file containing multiple targets 
described in Sec.~\ref{mtar}. The first stage was to produce calibrated target data for the two targets in the file. The second stage was to run 
the processing on the single source file. For imaging we used a pixel size of 1 arcsecond and an imagesize of 10000 pixels.
A total of 8 self-calibration loops with first 4 phase-only followed by 4 amplitude and phase calibration. 
For the legacy GMRT data a single pipeline run was used as described in Sec.~\ref{std}. We used a pixel size of 
1 arcsecond and an image size of 9000. The {same }self-calibration loops as for the uGMRT 
data were used. 
The central regions of the images produced using CAPTURE are presented in Fig.~\ref{fig:ugmrt} and Fig.~\ref{fig:legacy}.

\begin{figure}
    \centering
    \includegraphics[trim = {0.0cm 0.0cm 0.0cm 0.0cm}, clip,height=12cm]{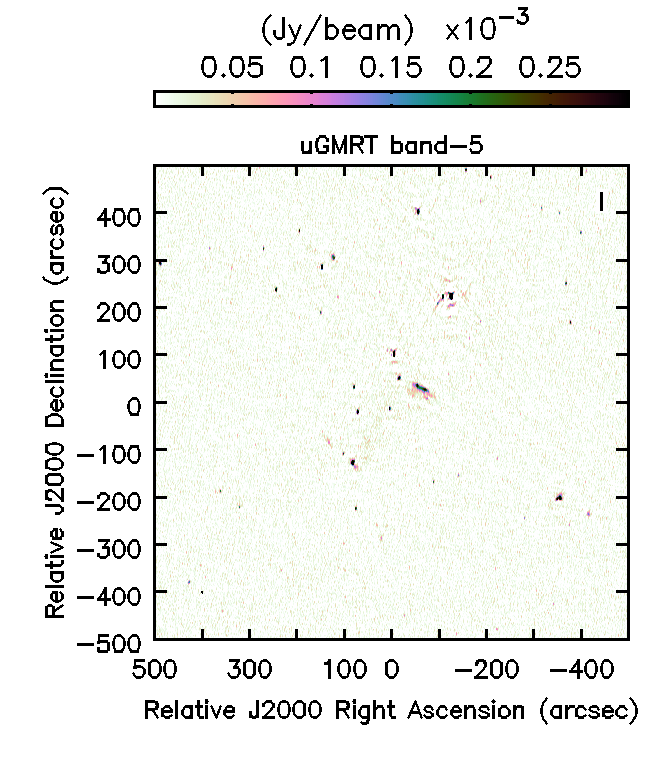}
    \caption{A uGMRT band-5 image with rms 17 $\mu$Jy beam$^{-1}$ produced using CAPTURE. The image has a synthesized beam of $5.6''\times2.1''$, with a position angle $-1.1^{\circ}$. 
    The central frequency of the image is 1274 MHz and an effective bandwidth of 332 MHz was used. The on-source time was $2$ hours. The field is located at declination $-49^{\circ}$.}
    \label{fig:b5ruta}
\end{figure}

\begin{figure}
    \centering
        \includegraphics[height=9cm]{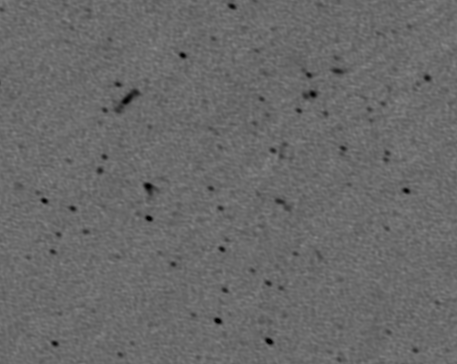}
    \caption{A uGMRT band-4 image with rms 6.7 $\mu$Jy beam$^{-1}$ produced using CAPTURE. The image has a synthesized beam of $5.0''\times3.5''$ and a size of $6'\times7.5'$ is displayed in grey 
    scale ranging between -10 and 500 $\mu$Jy beam$^{-1}$. The central frequency of the image is 690 MHz and an effective bandwidth of 250 MHz was used. The on-source time was $\sim5$ hours.}
    \label{fig:b4ich}
\end{figure}

Additional examples of the images produced using CAPTURE are shown in Fig.~\ref{fig:b5ruta} (band-5) and Fig.~\ref{fig:b4ich} (band-4).

\begin{table}
\caption{Summary of observations.}
\label{obs}       
\begin{tabular}{cccccc}
\hline\noalign{\smallskip}
Prop. code & Obs. Date &Duration& Freq. & rms & beam   \\
           &           & (min)  & MHz   & (mJy beam$^{-1}$) & ($''\times''$, PA)   \\

\noalign{\smallskip}\hline\noalign{\smallskip}
$34\_059$ & 22 Jul. 2018 &180& 300 - 500   & 0.043&$7.4\times5.5$, $76.5^{\circ}$\\

$31\_017$ & 20 Nov. 2016  &330&608  & 0.028 & $6\times4$, $67.5^{\circ}$ \\
\noalign{\smallskip}\hline
\end{tabular}
\end{table}

\section{Primary beam correction: making images science-ready}\label{pb}
In order to use the images for scientific analysis, they need to be corrected for the primary beam gain. CAPTURE uses the task tclean for imaging without the 
correction for the primary beam incorporated (the A-projection algorithm, \citet{2013ApJ...770...91B}) as the primary beam models for the GMRT as required by these algorithms 
are not yet available. 

Due to the wide bandwidth of the uGMRT, the primary beam changes across the frequency band significantly and needs to be accounted for in the correction.
 The CASA task ``widebandpbcor'' is the wideband primary beam correction task for the VLA. The CASA task ``wbpbgmrt''\footnote{\url{https://github.com/ruta-k/uGMRTprimarybeam}} is 
a modification of ``widebandpbcor'' to apply the corrections for the GMRT primary beam shapes. Currently wide-band primary beam model in the form of a polynomial approximation 
is available for the uGMRT\footnote{uGMRT primary beam polynomial approximation, Chengalur, J and Lal, D. V. private communication.}. This task requires the measurement set used for imaging, the image and a range of frequencies 
(channel numbers) within the spectral window provided by the user. 

For the legacy GMRT, the primary beam correction can be carried out in two steps. First step is to produce the image of the GMRT primary beam based on the polynomial approximation 
of the beam with the peak at the phase center of the target field. For this step the task ``gmrtpb'' has been developed\footnote{\url{https://github.com/ruta-k/GMRTpbcor}}. The image of the primary beam 
created by this task then should be used in the task ``impbcor'' for the correction to be applied to the image.

\section{Special cases of further user control}\label{spcases}
There are portions of CAPTURE where it can be easily modified to accommodate a few special cases of the data. 

The source names in any given file are compared with the standard names to classify them into calibrators and targets. We use the list of 
names of secondary calibrators in J2000 format from the VLA calibrator list (\url{https://science.nrao.edu/facilities/vla/observing/callist}). 
If the file contains calibrator names that are diffrent than these, then an easy work around is to find the names of the secondary calibrators and 
append it to the text file ``vla-cals.list''.

In the imaging and self-calibration part of the pipeline, the number of iterations (niter parameter in the task tclean) at the start is ``niter\_start'' 
and the maximum number of allowed iterations is hard coded to 200000. The user may change 
the parameter ``niter\_start'' in config\_capture.ini to a lower value such as 200 or 500 if the field is crowded with bright sources. The choice of a lower value will 
use less number of clean components in the model and results in a less chance of picking up artefacts in the model during the initial self-calibration iterations.

The choice of auto-masking in tclean controls the masking and if the user has expertise in tuning the sub-parameters in tclean, it can be done 
in the corresponding function in the file ``ugfunctions.py''. 
Auto-masking and myniterend prevent the tclean from diverging and in turn the parameter ``mJythreshold'' is rendered less critical. A value set 
equal to that of the expected final rms noise has worked well in our test cases. A high value of ``mJythreshold'' will result in premature stopping of tclean before cleaning 
the faint sources and therefore is not recommended.

\section{Limitations of CAPTURE}\label{limits}

Our approach in this pipeline has been to have no external dependency of software except CASA to make the pipeline usable on any machine where 
the user can run CASA. Thus use of external flaggers such as the AOFlagger \citep{2010ascl.soft10017O} are not implemented in the pipeline. 
However given the tool-kit like design of the pipeline, the user can use external flagging at intermediate steps 
and use the pipeline for further analysis. 

It is recommended to have disk space equivalent to at least six times that of the raw data size for running CAPTURE from 
end-to-end. The original lta, FITS and MS files are retained and their deletion is left to the user's discretion. 
The images with all the extensions as produced by tclean are also retained. The intermediate visibilities from self-calibration, except that 
from the final iteration, are deleted. 

At the GMRT, data can be recorded in the Indian polar (total intensity) or the Full Polar modes. In the default Indian Polar mode, only the RR and LL 
correlations are present. CAPTURE has been written for the Indian Polar mode data, however it has been run on data where RR, LL, RL and LR correlations 
were present. CAPTURE does not perform polarization calibration and thus the cross polarization data will essentially remain unused though the 
stokes I images will be produced. A version of CAPTURE incorporating polarization calibration is being developed.

For the uGMRT data presented here it took 4.3 days to complete 8 self-calibration iterations on a machine with 32 GB RAM and 8 cores. 
Since the user can tune the number of self-calibration loops and the wprojection planes, the run time can change considerably based on these choices.
CAPTURE run time is limited by the speeds of CASA tasks and python and thus we only provide numbers for the speed as seen in the CASA logger and rigorous speed tests 
are beyond the scope of this work.  
The step that takes most of the running time in CAPTURE is the task tclean. Users can modify the choice of ``$nwprojpl=-1$'' to 128 and ``imsize\_pix'' 
to a lower value at the cost of compromising on the w-term correction and the deconvolution of sources outside the area covered in 
the given image size to make the imaging run faster. 
The pipeline is currently being tested with the use of ``WSclean'' \citep{2014MNRAS.444..606O} to improve upon the 
speed of the imaging step.

{The CAPTURE pipeline presented here does not take into account the direction dependent effects such as due to the ionosphere and the primary beam.} 

\section{Conclusions}\label{conclu}
We have presented ``CAPTURE'' - CAsa Pipeline-cum-Toolkit for uGMRT Data REduction, a python and CASA based pipeline for the analysis of data obtained from the uGMRT at bands 3, 4 and 5 
and the legacy GMRT data. CAPTURE has no external dependency except that of CASA and the lta to FITS conversion pre-compiled codes for the uGMRT 
data. It can be used to produce self-calibrated images from the data in a fully automatic mode and can also be run in steps to allow users to inspect the data 
at various stages of the analysis. We have shown examples of running CAPTURE on uGMRT band-3 data and obtained an rms noise of $43\,\mu$ Jy beam$^{-1}$. CAPTURE can also work with 
legacy GMRT data and in the example shown here with 608 MHz data we obtained an image with rms noise of $ 28\,\mu$Jy beam$^{-1}$. {Additional two examples of images using CAPTURE 
at band 4 with an rms of $6.7 \mu$Jy beam$^{-1}$ and at band-5 with an rms $17 \mu$Jy beam$^{-1}$ are also shown.} In all these cases 
the rms is within the ballpark value expected given the limitations of the GMRT. The primary beam correction for the images is not part of CAPTURE and 
the tasks for carrying it { out} are made separately available on github.
CAPTURE is publicly available (\url{https://github.com/ruta-k/uGMRT-pipeline}, release v1.0.0) and is being developed further to improve on the performance 
and additional features such as polarization calibration {and other imaging options}.  

\begin{acknowledgements}
RK acknowledges the support from the DST-INSPIRE Faculty Award of the Government of India.
We acknowledge the support of the Department of Atomic Energy, Government of
India, under project no. 12-R\&D-TFR-5.02-0700. We thank the staff of the Giant Metrewave Radio Telescope who have made the observations possible.
The GMRT is run by the National Centre for Radio 
Astrophysics of the Tata Institute of Fundamental Research. We thank J. Chengalur and D. V. Lal for the polynomial approximation for the uGMRT primary beams. 
We thank Biny Sebastian for her work on the wideband primary beam correction  
for the uGMRT.
We also thank our colleagues who have tried CAPTURE and provided constructive feedback. 
\end{acknowledgements}

%
%

\bibliographystyle{spbasic}      
\bibliography{ruta_all_1.bib} 

%
%

\end{document}